\documentclass[aps,twocolumn,prl,showpacs,amsmath,amssymb,10pt]{revtex4-2}
\usepackage{graphicx}
\usepackage{xcolor}

\usepackage{float}
\usepackage{scrextend}
\usepackage{manfnt,pifont}
\usepackage{hyperref}
\usepackage{url}

\usepackage[normalem]{ulem}

\usepackage[outline]{contour}
\usepackage{textgreek}

\usepackage{tikz}
\usetikzlibrary{shapes}

\usepackage{CJK}
\usepackage{enumitem}
\setlist[itemize]{leftmargin=*}

\newcommand\wh\widehat

\newcommand\ri{{\mathrm{i}}}
\newcommand\re{{\mathrm{e}}}

\providecommand{\ee}{\mathrm{e}}
\providecommand{\ket}[1]{|#1\rangle}
\providecommand{\bra}[1]{\langle #1|}
\providecommand{\braket}[2]{\langle #1|#2\rangle}

\makeatletter
\DeclareFontFamily{OMX}{MnSymbolE}{}
\DeclareSymbolFont{MnLargeSymbols}{OMX}{MnSymbolE}{m}{n}
\SetSymbolFont{MnLargeSymbols}{bold}{OMX}{MnSymbolE}{b}{n}
\DeclareFontShape{OMX}{MnSymbolE}{m}{n}{
    <-6>  MnSymbolE5
   <6-7>  MnSymbolE6
   <7-8>  MnSymbolE7
   <8-9>  MnSymbolE8
   <9-10> MnSymbolE9
  <10-12> MnSymbolE10
  <12->   MnSymbolE-Bold12
}{}
\DeclareFontShape{OMX}{MnSymbolE}{b}{n}{
    <-6>  MnSymbolE-Bold5
   <6-7>  MnSymbolE-Bold6
   <7-8>  MnSymbolE-Bold7
   <8-9>  MnSymbolE-Bold8
   <9-10> MnSymbolE-Bold9
  <10-12> MnSymbolE-Bold10
  <12->   MnSymbolE12
}{}

\let\llangle\@undefined
\let\rrangle\@undefined
\DeclareMathDelimiter{\llangle}{\mathopen}
                     {MnLargeSymbols}{'164}{MnLargeSymbols}{'164}
\DeclareMathDelimiter{\rrangle}{\mathclose}
                     {MnLargeSymbols}{'171}{MnLargeSymbols}{'171}
\makeatother

\usepackage{eso-pic}
\newcommand{\reportnum}[2]{
  \AddToShipoutPictureBG*{
    \AtPageUpperLeft{
      \hspace{0.75\paperwidth}
      \raisebox{#1\baselineskip}{
        \makebox[0pt][l]{\textnormal{#2}}
  }}}
}

\begin{document}

\reportnum{-3}{USTC-ICTS/PCFT-26-30}

\title{Gate Parameter Lee-Yang Zeros and Dynamical Phases in Quantum Circuits}

\author{Yunfeng Jiang$^{a,c}$}
\thanks{Corresponding author: jinagyf2008@seu.edu.cn}

\author{Chang Liu$^{a}$}

\author{Yu Wu$^{b}$}

\author{Yang Zhang$^{b,c,d}$}
\thanks{Corresponding author: yzhphy@ustc.edu.cn}

\affiliation{$^a$School of Physics \& Shing-Tung Yau Center, Southeast University, Nanjing 211189, P. R. China}
\affiliation{$^b$Interdisciplinary Center for Theoretical Study, University of Science and Technology of China,
Hefei, Anhui 230026, China}
\affiliation{$^c$Peng Huanwu Center for Fundamental Theory, Hefei, Anhui 230026, China}
\affiliation{$^d$Center for High Energy Physics, Peking University, Beijing 100871, People's Republic of China}

\begin{abstract}
We propose gate-parameter Lee-Yang zeros of Loschmidt amplitudes as probes of dynamical phases in finite quantum circuits. We study Floquet circuits constructed from two-qubit fSim gates with identical parameters, for which the Loschmidt amplitude becomes a rational function of the gate parameters after a suitable change of variables. At fixed system size and large circuit depth, the zeros in one complexified gate parameter, with the other held fixed, condense onto limiting curves. In contrast to conventional Loschmidt or Fisher zeros in complex time, these zeros live directly in the complex plane of a tunable gate parameter. We show that the limiting set has two origins: a state-dependent component controlled by overlaps with Floquet eigenstates, and a universal component fixed by the Floquet spectrum. As the remaining gate parameter is varied, the universal zero set reorganizes abruptly, providing a finite-qubit diagnostic of a dynamical phase transition. We demonstrate this behavior in a Bethe ansatz solvable brickwork circuit and in longer-range fSim circuits outside this solvable structure. The mechanism follows from the Beraha-Kahane-Weiss theorem together with local unitarity, and is therefore spectral rather than a special consequence of integrability.

\end{abstract}

\maketitle

\noindent{\bf Introduction.}  Quantum circuit models serve as a natural framework for quantum computation \cite{NielsenChuang2010} and for the digital simulation of quantum many-body dynamics \cite{Feynman1982QS,Lloyd1996QS}. These models are now realizable in programmable quantum processors \cite{Arute2019QS}. Although a fully fault-tolerant quantum computer remains unavailable, quantum circuits in the noisy intermediate-scale quantum (NISQ) era have already found significant applications in quantum many-body systems, driving substantial theoretical and experimental progress in recent years. They have been employed to investigate nonequilibrium many-body phenomena, including entanglement growth \cite{Nahum2017RUC,Fisher2023RQC,Zhou:2018myl}, operator spreading \cite{vonKeyserlingk2018Hydro,Nahum:2017yvy,Rakovszky:2017qit}, quantum transport \cite{Vanicat:2018Itl,Vernier2023,morvan2022formation}, and dynamical phases beyond equilibrium \cite{Fisher2023RQC,2016PhRvL.116y0401K,Else:2016yfq}. Circuit models have also emerged as a theoretical testbed for monitored dynamics \cite{Li2018MIPT,Chan2019MIPT,Skinner2019MIPT,GoogleQAI2023MIPT,Hoke:2023nmt}. Exactly solvable circuit families, such as dual-unitary circuits \cite{Bertini2019DualUnitary,Piroli2020DualUnitary} and Yang-Baxter integrable Trotterizations \cite{Vanicat:2018Itl,Aleiner2021,Vernier2023,Miao:2023mcr,Miao:2022dau} offer complementary platforms where analytic results can be obtained.

A central goal of many-body physics is to classify phases, since phase structure provides a robust organizing principle for otherwise complicated microscopic dynamics. In quantum circuits this question takes a new form. The basic object is not a static Hamiltonian or a ground state, but a discrete time evolution operator assembled from experimentally tunable gates. Different choices of gate parameters can lead to distinct dynamical regimes. Since current quantum processors operate at finite system size and finite circuit depth, it is important to develop diagnostics of such dynamical phases that do not rely on the thermodynamic limit. At the same time, the natural observable of quantum circuits are transition amplitudes, or Loschmidt amplitudes, which can be measured in circuit experiments and encode both the Floquet spectrum and the information of the initial state. This raises the question of whether the structure of finite circuit transition amplitudes can reveal sharp dynamical regimes.\par

We answer this question by proposing gate-parameter Lee-Yang zeros of Loschmidt amplitudes as
probes of distinct dynamical regimes. For a fixed circuit architecture and initial state, the
Loschmidt amplitude can be viewed as a cylinder partition function. Its zeros, after analytic
continuation of gate parameters, are therefore natural analogs of Lee-Yang zeros in statistical
mechanics \cite{YangLee1952,LeeYang1952,Bena2005LY}. This construction is closely related to,
but different from, the conventional Loschmidt-zero approach to dynamical quantum phase
transitions. In the latter, one fixes the Hamiltonian or Floquet operator and studies zeros in
complex time. Real-time nonanalyticities emerge in the thermodynamic limit
\cite{HeylM2013DQP,Karrasch2013DPT,Vajna2014DQPT,Budich2016DQPT,heyl2018DQP,GuSun2026}. Here,
by contrast, the circuit depth remains the discrete evolution parameter, while one tunable gate
parameter is complexified. Thus the relevant zeros live in a complex gate-parameter plane rather
than in the complex-time plane.\par

At fixed system size and large circuit depth, these zeros condense onto limiting curves. We show
that the limiting set has two origins: a state-dependent component and a universal component determined by the Floquet spectrum. As the remaining
gate parameter is varied through a critical value, the universal zero set reorganizes abruptly,
providing a finite-size diagnostic of a dynamical phase transition.\par

We demonstrate this mechanism in quantum circuits built from the two-qubit fSim gate, a
natural building block for superconducting quantum processors. We first analyze the brickwork
Floquet circuit, which is Bethe-ansatz solvable and
allows the Loschmidt amplitude to be computed exactly. We then study two longer-range circuit
architectures that are beyond Bethe ansatz solvable structure. The same reorganization of
the universal zero set is observed in all cases. Its origin can be understood from the
Beraha-Kahane-Weiss theorem together with local unitarity of the fSim gate.  The universal loci are therefore fixed by local unitarity for any Floquet operator assembled
from identical fSim gates, while the remaining branches retain model- and state-dependent
information.

\vspace{0.5cm}

\noindent{\bf The models.}
We consider several circuit models constructed from the two-qubit fSim gate \cite{Kivlichan:2018bqq,GoogleAIQuantum:2020hhm}
\begin{align}
\text{fSim}(\alpha,\phi)=\begin{pmatrix}
1&0&0&0\\
0&\cos\alpha&\ri\sin\alpha&0\\
0&\ri\sin\alpha&\cos\alpha&0\\
0&0&0&\re^{2\ri\phi}
\end{pmatrix}\,,
\end{align}
which depends on two parameters, $\alpha$ and $\phi$. It is convenient to work with a gauge-transformed two-qubit gate $U_{ij}(\alpha,\phi)$ acting on sites $i,j$ \cite{Aleiner2021} which is related to the fSim gate as
\begin{align}
\text{fSim}_{i,j}(\alpha,\phi)
=\re^{\ri\phi}\re^{-\ri\phi(\sigma_i^z+\sigma_j^z)/2}U_{ij}(\alpha,\phi)\,.
\end{align}

\vspace{0.3cm}
\noindent{\it - The model.} Using the fSim gate, or its gauge-transformed version $U_{ij}$  as a building block, one can construct quantum circuit models with rich structures. For a system of $L$ qubits, we define a discrete time-evolution Floquet operator $\hat{\mathcal{F}}$ and choose an initial state $|\Psi\rangle$. The quantity of interest is the following Loschmidt amplitude
\begin{align}
\mathcal{D}_{\Psi}(n)=\langle\Psi|\hat{\mathcal{F}}^n|\Psi\rangle\,.
\end{align}
We will study several different Floquet operators $\hat{\mathcal{F}}$. A well-known model is the brickwork circuit, whose Floquet operator reads
\begin{align}
\hat{\mathcal{F}}_{\text{bw}}=\mathcal{U}_e\mathcal{U}_o,
\end{align}
with
\begin{align}
\mathcal{U}_o=U_{12}U_{34}\cdots U_{L-1,L},\qquad
\mathcal{U}_e=U_{23}U_{45}\cdots U_{L,1},
\end{align}
where $L=2l$ is even.
%\begin{figure}[tbp]
%\centering
%\includegraphics[scale=0.55]{QC.pdf}
%\caption{Construction of the brickwork Floquet circuit.}
%\label{fig:QC}
%\end{figure}
This model is Bethe-ansatz solvable \cite{Vanicat:2018Itl,Aleiner2021}. One important observation is that the gauge-transformed gate $U_{ij}(\alpha,\phi)$ can be identified directly with the six-vertex $\check{R}$-matrix \cite{Baxter1982,Korepin1993}. Under the change of variables
\begin{align}
\label{eq:relatePara}
\tan\alpha=-\ri\frac{\sinh \xi}{\sinh\eta},\quad e^{-2\ri\phi}=-\frac{\sinh(\xi-\eta)}{\sinh(\xi+\eta)}\,.
\end{align}
we write $U_{12}(\alpha,\phi)=\check{R}_{12}(\xi,\eta)$. Omitting the anisotropy $\eta$ from the notation, the $\check{R}$-matrix satisfies the Yang-Baxter equation
\begin{align*}
\check{R}_{12}(u)\check{R}_{23}(u+v)\check{R}_{12}(v)=
\check{R}_{23}(v)\check{R}_{12}(u+v)\check{R}_{23}(u)\, .
\end{align*}
This is the origin of the Yang-Baxter integrability of the model, which allows us to obtain the spectrum of $\hat{\mathcal{F}}_{\text{bw}}$ by Bethe ansatz and to compute the Loschmidt amplitude much more efficiently.\par

Not all Floquet operators, however, are Bethe-ansatz solvable. To go beyond the brickwork model, we introduce the range-$r$ operator
\begin{align}
\mathcal{V}_r=\prod_{j=1}^L U_{j,j+r}\,.
\end{align}
We consider two additional models defined by the Floquet operators
\begin{align}
\hat{\mathcal{F}}_{\text{II}}=\mathcal{V}_2 \mathcal{U}_o,\qquad \hat{\mathcal{F}}_{\text{III}}=\mathcal{V}_3\mathcal{V}_2\mathcal{U}_o\,.
\end{align}
These models are defined rather arbitrarily in order to break the explicit solvability of the brickwork model and to illustrate the universality of our observations.

%\vspace{0.3cm}
%\noindent{\it - Bethe ansatz.} The gate conserves total $S^z$, so $\mathcal{U}$ decomposes into
%sectors with a fixed number $M$ of magnons. In each sector, Bethe ansatz eigenstates are denoted
%by $|\mathbf{u}_M\rangle$, where $\mathbf{u}_M=\{u_1,\ldots,u_M\}$ are Bethe roots. They satisfy
%\begin{align}
%\mathcal{U}|\mathbf{u}_M\rangle=\tau(\mathbf{u}_M)|\mathbf{u}_M\rangle 
%\end{align}
%with
%\begin{align}
%\tau(\mathbf{u}_M)=\prod_{k=1}^M\frac{\sinh(u_k-\xi/2+\eta)\sinh(u_k+\xi/2)}{\sinh(u_k+\xi/2+\eta)\sinh(u_k-\xi/2)}
%\end{align}
%provided the roots $\mathbf{u}_M$ solve the Bethe ansatz equations (BAE), whose form can be found in SM. 
%\begin{align}
%\left(\frac{\sinh(u_k-\xi/2+\eta)\sinh(u_k+\xi/2+\eta)}{\sinh(u_k-\xi/2)\sinh(u_k+\xi/2)}\right)^l
%=\prod_{k\ne j}^M\frac{\sinh(u_k-u_j+\eta)}{\sinh(u_k-u_j-\eta)}
%\end{align}

\vspace{0.3cm}
\noindent{\it - Initial states.} To compute the Loschmidt amplitude we must also specify the initial state. We consider four representative choices:
\begin{itemize}
\item Domain wall state
\begin{align}
|\text{DW}_M\rangle=|\underbrace{1\ldots 1}_M0\ldots 0\rangle
\end{align}
\item Dimer state
\begin{align}
|\Psi_{\text{dimer}}\rangle=\prod_{j=1}^{l}(\sigma_{2j-1}^x-\sigma_{2j}^x)|00\ldots 0\rangle
\end{align}
\item Ne\'el state
\begin{align}
|\Psi_{\text{Ne\'el}}\rangle=\prod_{j=1}^l\sigma_{2j}^x|00\ldots0\rangle
\end{align}
\item Crosscap state
\begin{align}
|\Psi_{\text{crosscap}}\rangle=\prod_{j=1}^l(1+\sigma_j^x\sigma_{j+l}^x)|00\ldots 0\rangle
\end{align}
\end{itemize}

\vspace{0.3cm}
\noindent{\it - Computational method.} In what follows we will need $\mathcal{D}_{\Psi}(n)$ for different $n$ with our main interest in the regime of large $n$. For the Bethe ansatz solvable brickwork model,  the Loschmidt amplitude can be computed exactly by using integrability \cite{Gaudin1983Bethe,korepin1982BEF,Slavnov1989ScalarProducts,Izergin1987partition,Izergin1992,Kostov2012IPBS,MosselCaux2010DomainWall,Brockmann2014NeelXXZ,Pozsgay2014ProductOverlaps,JiangPozsgay2020ExactOverlaps,CaetanoKomatsu2022Crosscap} and computational algebraic geometry \cite{Jiang2018AG,Jacobsen2019AG,Bajnok2020AG,Hutsalyuk2024sawAG}. More details can be found in the supplementary material.\par

For $\hat{\mathcal{F}}_{\text{II}}$ and $\hat{\mathcal{F}}_{\text{III}}$, which are not Bethe ansatz solvable, we evaluate the Loschmidt amplitude by a different approach. We evaluate the action of the Floquet operators on the state in a recursive manner, implementing the action of the Floquet operator as a sparse matrix, and finally projecting out the initial state to obtain the amplitude.  With proper implementation, this method can be made rather efficient. More details are delegated to the supplementary material.

%Since every two-site gate conserves total $S^z$, we work in the fixed-$M$ computational basis of dimension $\binom{L}{M}$. Starting from $|\Psi\rangle$, we apply the gates bond by bond for $n$ Floquet periods and update the basis-state amplitudes by exact sparse-polynomial dynamic programming. Projecting the evolved state onto $\langle\Psi|$ then gives $\mathcal{D}_{\Psi}(n)$ as an exact rational function of $q$ and $x$.
We focus on systems  at  finite sizes, the large-depth regime $n\sim100$ shows clear pattern for the distribution of zeros, which typically condense on some limiting curves. We focus on systems with $L=6\sim 14$ qubits. Even for these rather small number of qubits, the large-depth regime shows clear signatures of the transition. In fact, as we will see, the shape of the universal limiting curves do not depend on the number of qubits.

\vspace{0.5cm}
\noindent{\bf Gate-parameter Lee-Yang zeros.} To analyze the Loschmidt amplitude, it is convenient to introduce the variables $q=\re^{\eta}$ and $x=\re^{\xi/2}$. After this change of variables, the Loschmidt amplitude becomes a rational function of $q$ and $x$ and can be written as
\begin{align}
\mathcal{D}_{\Psi}(n)=\frac{P_{\Psi,n}(q,x)}{Q_{\Psi,n}(q,x)}\, ,
\end{align}
where $P_{\Psi,n}$ and $Q_{\Psi,n}$ are coprime polynomials. The zeros of $P_{\Psi,n}$ are zeros of the Loschmidt amplitude. For the examples considered here, the denominator $Q_{\Psi,n}(q,x)$ factorizes into an initial-state-dependent common factor and a power of $(1-q^2x^4)$, with an exponent fixed by $n$, $L$, and $M$. We therefore study the zeros of
$P_{\Psi,n}(q_0,x)$ as a polynomial in the complex variable $x$ with $q=q_0$ fixed to be a constant. We call these zeros the \emph{gate-parameter Lee-Yang (GPLY) zeros} of the Loschmidt amplitude. Furthermore, we introduce the anisotropy parameter $\Delta=(q+q^{-1})/2$. Following the terminology of the XXZ spin chain, we call $|\Delta|<1$ and $|\Delta|>1$ the massless and massive regimes, respectively. 
%We approach the transition at $\Delta=1$ from $q=\re^{\ri\gamma}$ on the massless side and from real $q>1$ on the massive side. Visible patterns already appear for even relatively small $n\sim L$. For large $n\gg 1$, the zeros condense onto curves, which we call limiting curves. 

\vspace{0.5cm}
\noindent{\bf The brickwork model.}
We first illustrate the analysis for the brickwork model.

\begin{figure}[tbp]
\centering
\includegraphics[scale=0.3]{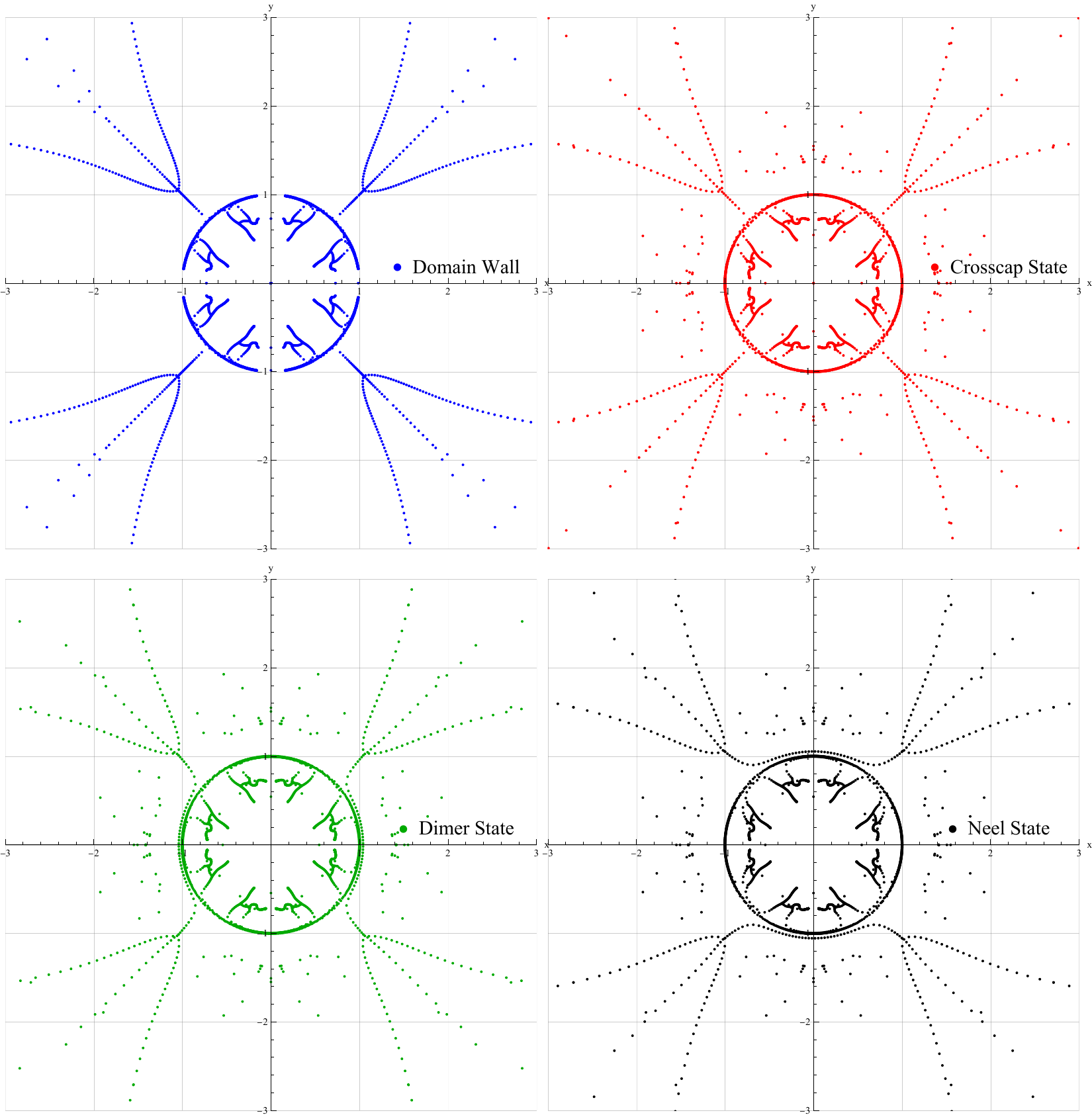}
\caption{GPLY zeros in the massive regime. The four plots show zeros of the reduced
numerator $P_{\Psi,n}(q_0,x)$ for the four indicated initial states. Here $q_0=2$
($\Delta=5/4$), $L=8$, $M=4$, and $n=200$.}
\label{fig:Massive}
\end{figure}

\vspace{0.3cm}
\noindent{\it - Massive regime.} In the massive regime, typical distributions of GPLY zeros in the complex $x$ plane for the four initial states are shown in Fig.~\ref{fig:Massive}. For different choices of $q_0$ within the regime, the distributions are similar.
The GPLY zeros exhibit very similar patterns for all four initial states, although some details remain state dependent. A common feature is that the limiting curves contain a unit circle. The zero distribution also exhibits a dihedral symmetry generated by a $\pi/2$ rotation and reflection about one axis.\par

In general, the limiting curve of the GPLY zeros fall into the following two classes: one that depends only on the Floquet operator $\hat{\mathcal{F}}_{\text{bw}}$ and the other that also depends on the initial state $|\Psi\rangle$. We call the first component universal, since it is independent of the choice of initial state. 

\begin{figure}[tbp]
\centering
\includegraphics[scale=0.3]{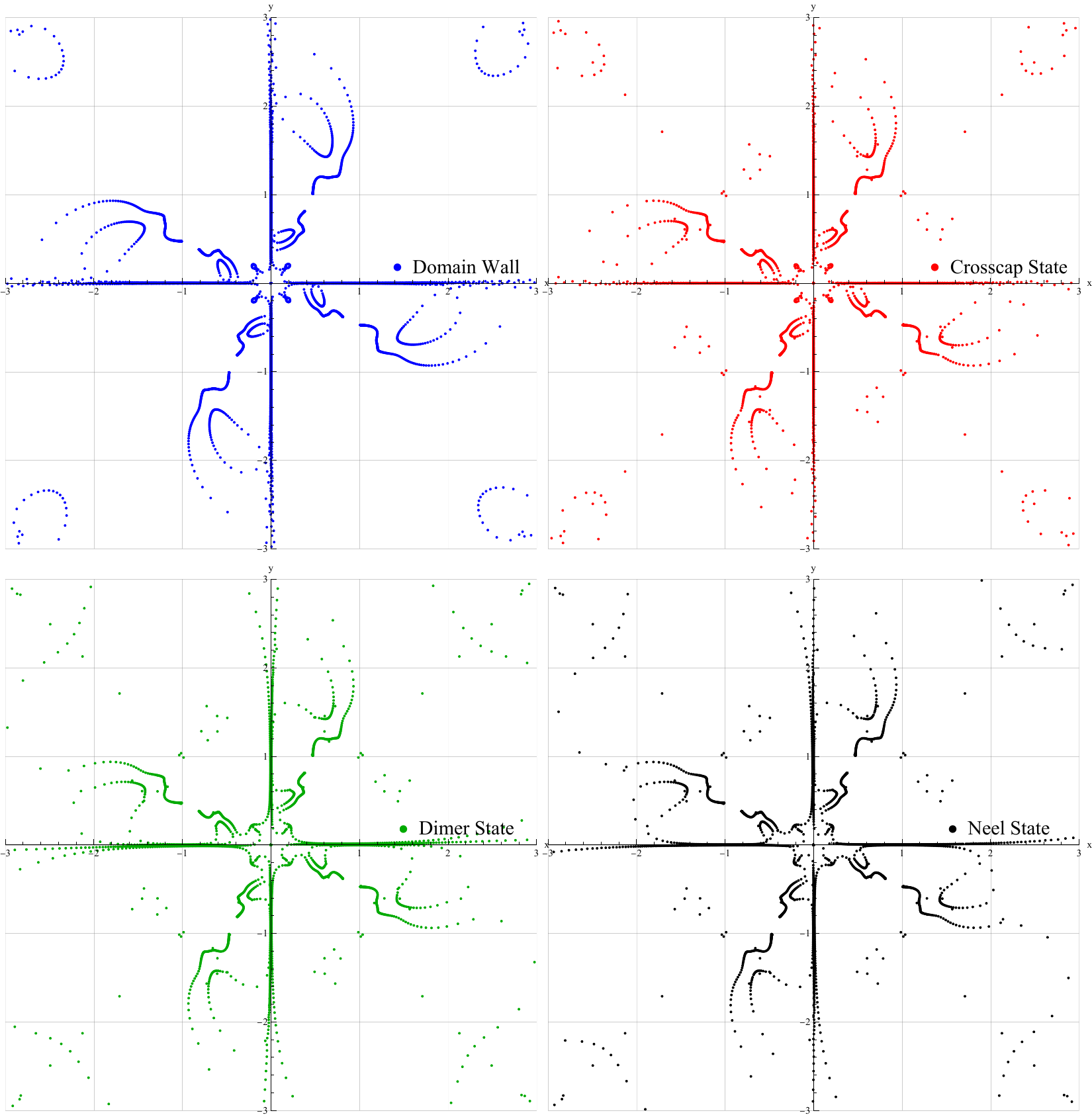}
\caption{GPLY zeros in the massless regime. The four plots show zeros of the reduced
numerator $P_{\Psi,n}(q_0,x)$ for the four indicated initial states. Here
$q_0=\frac{3}{5}+\frac{4}{5}\ri$ ($\Delta=3/5$), $L=8$, $M=4$, and $n=200$.}
\label{fig:Massless}
\end{figure}

\vspace{0.3cm}
\noindent{\it - Massless regime.} In the massless regime, typical distributions of GPLY
zeros in the complex $x$ plane for the four initial states are shown in Fig.~\ref{fig:Massless}.
\begin{figure*}[t]
\centering
\includegraphics[width=0.7\textwidth]{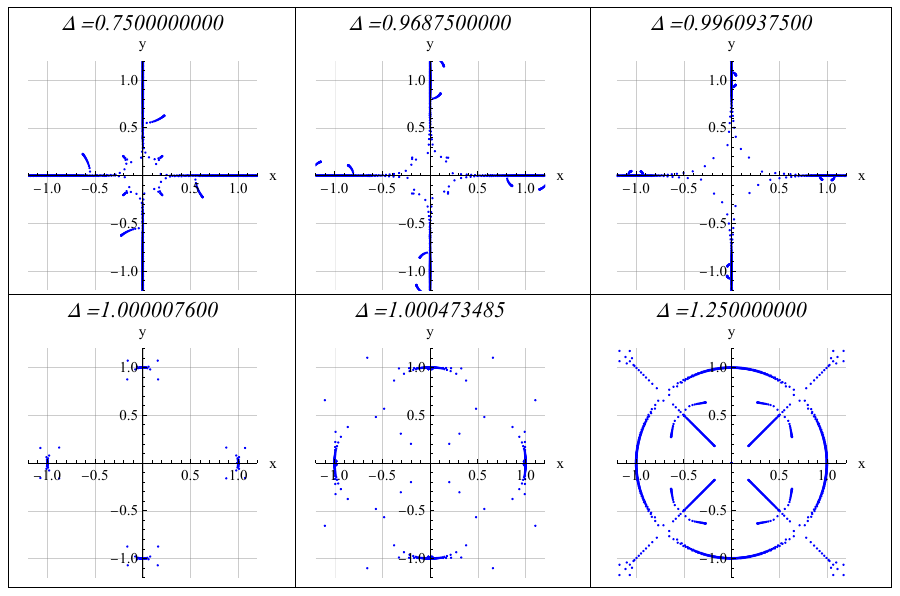}
\caption{Evolution of the GPLY zeros around $\Delta=1$. Here the initial state is the $L=8$
domain wall state $|\Psi\rangle=|11000000\rangle$, with $n=100$. Each panel contains 1592 zeros
counted with multiplicity.}
\label{fig:PT}
\end{figure*}
In the massless regime, the GPLY zeros for the four initial states again share a common
structure, but also with some different details that are state dependent. It is quite obvious that the distributions are qualitatively different
from those in the massive regime: the unit circle disappears, and in all cases there are branches of limiting curves that lie along the
real and imaginary axes. The zeros also form spiral-like patterns. They exhibit a $\mathbb{Z}_4$
symmetry, while the reflection symmetry present in the massive case is lost. Further tests show that all GPLY zeros in the regime $|\Delta|<1$ exhibit very similar structure.

\vspace{0.3cm}
\noindent{\it - Phase transition.} The comparison of the two regimes suggests a transition at $\Delta=1$. To see this more clearly, we track the GPLY zeros for the domain wall state with $L=8$ and $M=2$, as shown in Fig.~\ref{fig:PT}. For $\Delta<1$, the universal limiting
curves lie on the coordinate axes and develop spiral arms. As $\Delta$ approaches $1$, the zeros concentrate near $x=\pm1$ and $x=\pm\ri$. For $\Delta>1$, the distribution reorganizes and a circular condensation curve emerges. This reorganization underlies the abrupt change in the universal limiting curves.

\begin{figure}[htp]
 \centering
 \includegraphics[width=0.48\linewidth]{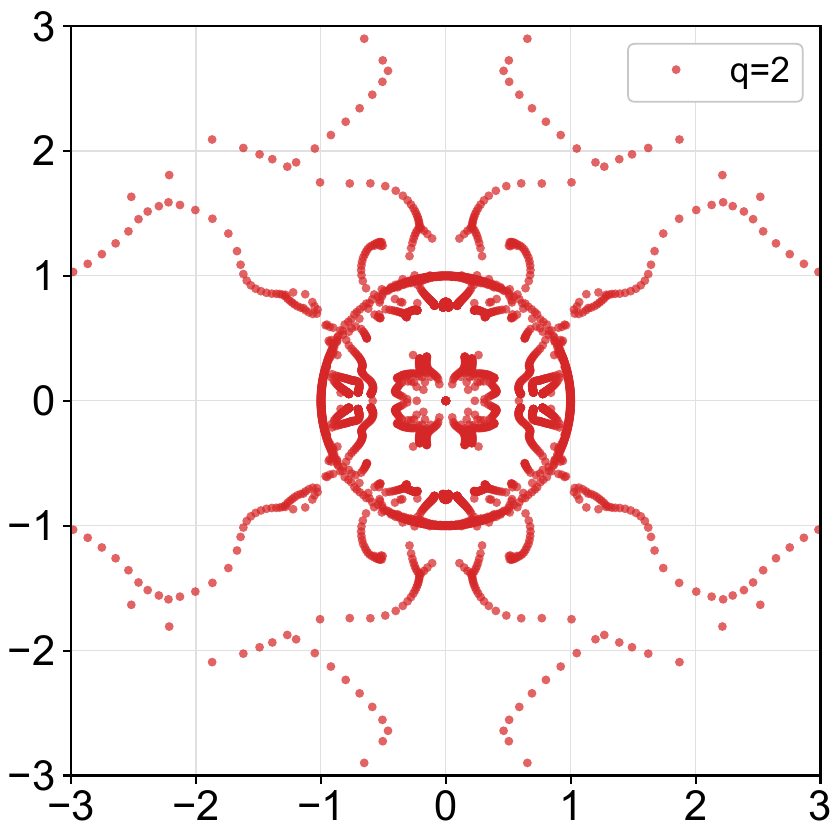}
 \hspace{0.001\linewidth}
 \includegraphics[width=0.48\linewidth]{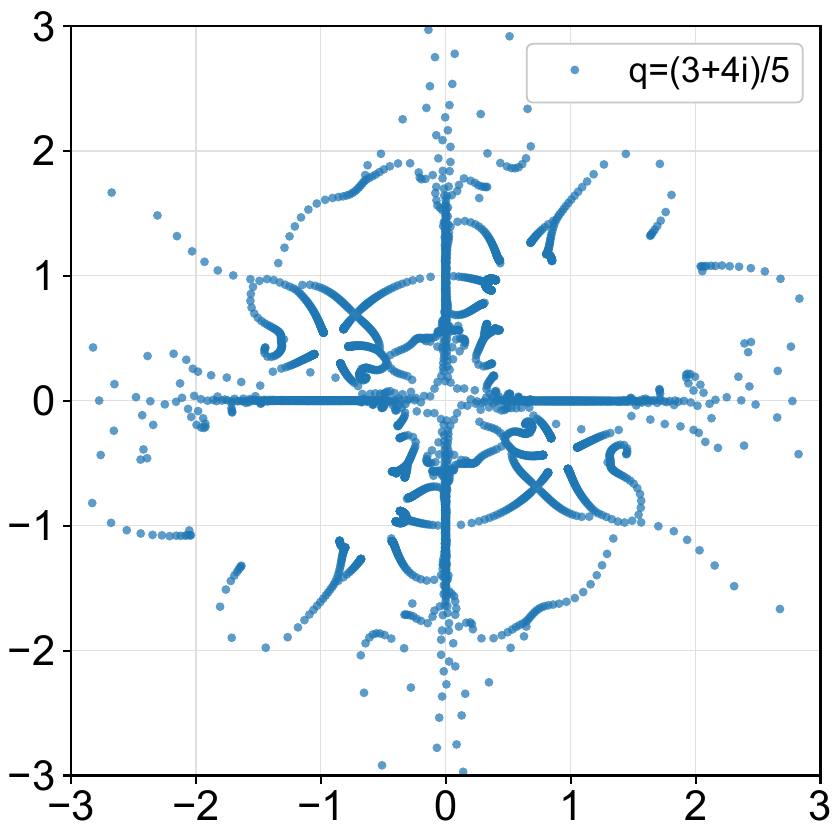}
 \caption{{GPLY-zero loci for the $\hat{\mathcal{F}}_{\text{II}}$ circuit in the massive regime (left) and massless regime (right).}}
 \label{sm:fig:range-2-zeros}
\end{figure}

\begin{figure}[htp]
 \centering
 \includegraphics[width=0.48\linewidth]{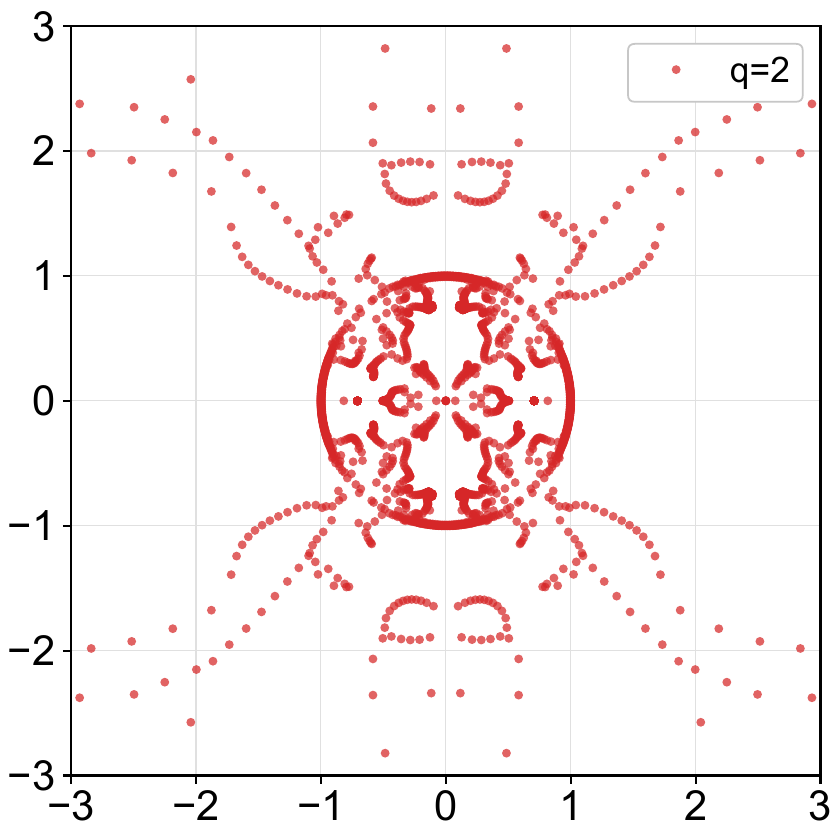}
 \hspace{0.001\linewidth}
 \includegraphics[width=0.48\linewidth]{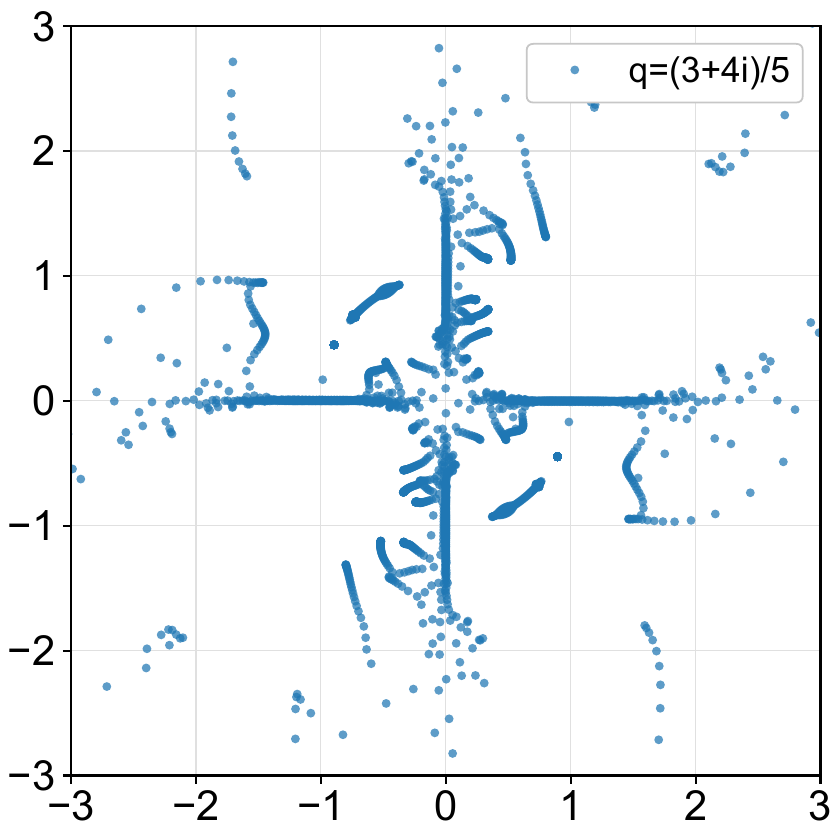}
 \caption{{GPLY-zero loci for the $\hat{\mathcal{F}}_{\text{III}}$  circuit in the massive regime (left) and massless regime (right).}}
 \label{sm:fig:range-3-zeros}
\end{figure}

\vspace{0.5cm}
\noindent{\bf General circuit models.} The phenomena persist for models that are not exactly solvable. We use the domain wall state with $L=8$ and $M=2$ as the initial state. Taking $n=100$, the GPLY zeros for $\hat{\mathcal{F}}_{\text{II}}$ and $\hat{\mathcal{F}}_{\text{III}}$ are shown in Fig.~\ref{sm:fig:range-2-zeros} and Fig.~\ref{sm:fig:range-3-zeros}, respectively.

We see that in these two models the loci of the GPLY zeros looks more complicated. Nevertheless, the unit circle and the condensation curves on the axes are still present. The patterns are again qualitatively different in the massive and massless regimes. As in the brickwork model, an abrupt change occurs at $\Delta=1$.

\vspace{0.5cm}
\noindent{\bf Universality.} The previous analysis shows that for both solvable and non-solvable models, and for different initial states, the limiting curves always contain a circle in the massive regime and line segments on the axes in the massless regime. An abrupt change of the patterns occurs at $\Delta=1$. We now explain the origin of this behavior.

The limiting curves can be analyzed using the Beraha-Kahane-Weiss theorem \cite{Beraha1978}. Consider a sequence of polynomials $P_n(x)$ of the form
\begin{align}
\label{eq:Pnexpand}
P_n(x)=\sum_{i=1}^k \alpha_i(x)\,\lambda_i(x)^n\,,
\end{align}
where $\lambda_i(x)$ are analytic functions of the complex variable $x$ on a chosen branch and $\alpha_i(x)$ are coefficient functions independent of $n$. The BKW theorem states that a point $z$ can be a limit point of zeros of $P_n$ through two mechanisms:
\begin{enumerate}
\item A single branch $\lambda_k(z)$ is strictly dominant at $z$, but its coefficient vanishes, $\alpha_k(z)=0$. We call these \emph{dominant curves}.
\item Two or more dominant branches are equimodular at $z$, the corresponding curves are called \emph{equimodular curves}.
\end{enumerate}
In the present context, let us denote the normalized eigenstates of the Floquet operator by $|i\rangle$ with eigenvalue $\tau_i(x)$. The Loschmidt amplitude can be written as
\begin{align}
\mathcal{D}_{\Psi}(n)=\langle\Psi|\hat{\mathcal{F}}^n|i\rangle\langle i|\Psi\rangle=\sum_i\tau_i(x)^n |\langle\Psi|i\rangle|^2
\end{align}
which takes the same form as in \eqref{eq:Pnexpand}. The dominant curves depend explicitly on $|\langle\Psi|i\rangle|^2$. The equimodular curve is spectral: it depends only on the Floquet eigenvalues and persists for all initial states whose overlaps with the eigenstates do not vanish identically. This second component is the universal part of the limiting curves.

Local unitarity provides a simple way to determine the universal curves. For real $\alpha$ and $\phi$ the two-site gate is unitary, \emph{i.e.}, $U^\dagger U=1$. Hence the Floquet operator $\hat{\mathcal{F}}$ is unitary and all its eigenvalues are unimodular. After analytic continuation to complex $x$, the loci where the parametrization \eqref{eq:relatePara} returns real $\alpha$ and $\phi$ give equimodular curves, which are precisely the universal limiting curves of GPLY zeros according to BKW. The reality of $\phi$ gives
\begin{align}
\label{eq:condition}
|\sinh(\xi-\eta)|=|\sinh(\xi+\eta)|\,.
\end{align}

\vspace{0.3cm}
\noindent{\it - Massive regime.}  In the massive regime $\eta\in\mathbb{R}$, set $\xi=m+\ri n$ with $m,n\in\mathbb{R}$. Substituting into Eq.~\eqref{eq:condition} gives $\sinh(2m)\sinh(2\eta)=0$. For $\eta\ne0$, this implies
$m=0$, so $\xi$ is purely imaginary and $|x|=|\re^{\xi/2}|=1$. The reality condition for
$\alpha$ is also satisfied. This explains the robust unit circle in Fig.~\ref{fig:Massive}.
The additional reflection and fourfold rotational symmetries follow from the invariances of the
parameter map under $\xi\to-\xi$, $\xi\to\xi+\ri\pi$, and complex
conjugation, which act on the $x$ plane as inversion, $x\to\ri x$, and reflection. These
symmetries generate the observed dihedral pattern. 

\vspace{0.3cm}
\noindent{\it - Massless regime.} In the massless regime $\eta=\ri\gamma$ with $0<\gamma<\pi$, we write $\xi=m+\ri n$. Eq.\eqref{eq:condition} gives $\sin(2n)\sin(2\gamma)=0$.
For generic $\gamma$, the reality of $\alpha$ further restricts the allowed unitarity locus to $n=0$ modulo $\pi$. Since $x=\re^{\xi/2}$, this means that $x$ is either real or purely imaginary. 
This explains the universal curves on the real and imaginary axes in Fig.~\ref{fig:Massless}. The shift $\xi\to\xi+\ri\pi$ maps the real axis to the imaginary axis in the $x$ plane, giving the observed $\mathbb{Z}_4$ rotational structure. The shift $\xi\to\xi+\ri\pi$, equivalently $x\to\ri x$, exchanges the real and imaginary axes and thus leaves their union invariant. This geometric invariance does not, however, imply an exact $\mathbb{Z}_4$ symmetry of the complete GPLY-zero set. 
 In our normalization, the local gate satisfies $U_{ij}(q,\ri x)=\sigma_i^z U_{ij}(q,x)\sigma_j^z$. For the alternating brickwork circuit, these endpoint rotations telescope into a global sublattice conjugation of $\hat{\mathcal F}$; for compatible initial states this gives $\mathcal{D}_{\Psi,n}(q,\ri x)\propto \mathcal{D}_{\Psi,n}(q,x)$, or equivalently a definite parity under $y=x^2\to-y$. For generic quantum circuit model, the endpoint rotations do not always telescope into a single global similarity transformation, and the two $y$-parity sectors can mix. Local unitarity therefore still selects the real and imaginary axes as equimodular loci, whereas the off-axis zeros and their densities need not be related by a quarter turn. The loss of an exact $\mathbb{Z}_4$ symmetry is thus schedule dependent and should not be attributed to nonintegrability alone.
Reflection symmetry is reduced relative to the massive case because the complex value of $q$ fixes an orientation in the analytically continued parameter space.\par

The same reasoning applies to general Floquet operator $\hat{\mathcal{F}}$ constructed from fSim gates with identical parameters. For such circuit models, the gate is unitary for real $\alpha$ and $\phi$, the eigenvalues of the Floquet operator are unimodular, and we therefore expect a circular equimodular condensation curve in the massive regime and condensation curves along the axes in the massless regime. An abrupt reorganization of the zeros occurs at $\Delta=1$. This demonstrates that the phase transition we observed is a generic feature for the quantum circuit models under consideration.

\begin{figure}[t]
\centering
\includegraphics[width=\columnwidth]{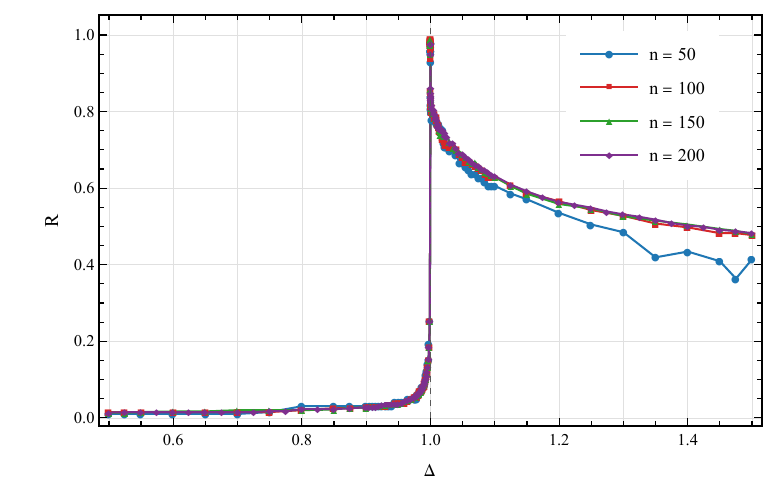}
\caption{Zero-density diagnostic near the universal curves. The plotted quantity
$R(\Delta;\epsilon,n)$ is the fraction of zeros within distance $\epsilon=10^{-2}$ of the
expected universal set: the coordinate axes for $\Delta<1$ and the unit circle for $\Delta>1$.}
\label{fig:R_delta}
\end{figure}

\vspace{0.3cm}
\noindent{\it - Phase transition and scaling.}
The transition can be quantified by the density of zeros near the universal curves. Let
$R(\Delta;\epsilon,n)$ denote the fraction of zeros lying within a distance $\epsilon$ of the
expected universal set: the unit circle in the massive regime and the coordinate axes in the
massless regime. One may then study the limits $n\to\infty$ followed by $\epsilon\to0$. For the brickwork model, at fixed
$\epsilon=10^{-2}$, Fig.~\ref{fig:R_delta} shows a rapid reorganization of this zero-density
diagnostic across $\Delta=1$. In the massive regime, fits for $n=100,150,200$ are consistent
with $R=1-c\epsilon^r$, with $c=0.47\pm0.2$ and $r=0.15\pm0.1$.
% Further numerical evidence and analytic details will be presented elsewhere \cite{LiuWuJiangZhangToAppear}.

%Equivalently, the rate function
%\begin{align}
%f_{\Psi}(x)=\lim_{n\to\infty}\frac{1}{n}\log|\mathcal{D}_{\Psi}(n)|
%\end{align}
%is expected to become nonanalytic across the limiting curves. These diagnostics turn the visual
%reorganization in Fig.~\ref{fig:PT} into a quantitative finite-size, long-time criterion for the
%dynamical transition. 

\vspace{0.5cm}
\noindent{\bf Discussions.} We have shown that gate-parameter Lee-Yang zeros of Loschmidt amplitudes provide a
sharp probe of dynamical phases in finite quantum circuits. In general, there are two types of limiting zero curves: universal curves fixed
by the Floquet spectrum and state-dependent curves fixed by overlaps with the initial state. For the brickwork model, the
universal curves reorganize at $\Delta=1$, producing a finite-size dynamical phase transition in
the large circuit depth limit.\par

The origin of the universal curves is spectral rather than state specific. By the BKW theorem, equimodular Floquet eigenvalues generate limiting zeros independently of the detailed overlap coefficients, provided the relevant overlaps do not vanish identically. Local unitarity gives a simple geometric characterization of these equimodular loci, explaining the unit circle in the massive regime and the coordinate axes in the massless regime.\par

It is useful to contrast this diagnostic with the standard dynamical quantum phase transition (DQPT) use of Loschmidt zeros. Conventional DQPTs fix the dynamics and study zeros in complex time, with real-time nonanalyticities emerging after a thermodynamic limit. Our construction fixes one circuit parameter and studies zeros in the other gate parameter, with the sharp structure emerging in the large-depth limit at fixed finite $L$. The two approaches are complementary slices of the same broader analytic object, the Loschmidt amplitude as a function of time/depth and circuit parameters. Both are governed by competition between spectral contributions, but they diagnose different projections of that competition.\par

The diagnostic is naturally suited to present-day quantum processors. Loschmidt amplitudes can
be accessed through interferometric protocols or related echo measurements, and Lee-Yang zeros
have already been inferred experimentally from quantum coherence and full counting statistics
\cite{Wei2012LY,Peng2015LY,Brandner2017EDD}. Partition-function zeros have also been proposed as
a route to many-body thermodynamics on quantum computers \cite{Francis2021PZQ}. Since the
transition studied here is formulated at fixed $L$ and $n$, it does not require
extrapolation to a thermodynamic limit.\par 

Several questions remain open. First, the finite-time
scaling of the zero density near $\Delta=1$ should be developed into a quantitative critical
diagnostic, it would be interesting to see whether the exponents are universal and can be
determined analytically. Second, in this work we have focused on the universal component of the
Lee-Yang zeros, this does not mean the state-dependent part is un-important. Instead, since they
encode rich information of the initial state, they can serve as useful probes for different kind of quantum states. A
systematic study of the state-dependent part of the GPLY zero and their potential applications
in quantum computation is a highly interesting task.

\vspace{0.5cm} 
\noindent{\bf Acknowledgements.} We thank Yuan Miao, Jue Hou, Xin Qian and Xiao Wang for interesting discussions related to this work, Yongqun Xu for guidance and assistance with the Linux terminal, and Jie Gu for support with the computing cluster. This is supported by the National Natural Science Foundation of China through Grant No. 12575078, 12575073 and 12247103.

\nocite{PalettaDuhPozsgayZadnik2025}
\bibliography{reference} 
\bibliographystyle{utphys}

\clearpage
\onecolumngrid

\begin{center}
{\large\bfseries Supplemental Material for\\[0.3em]
``Gate Parameter Lee-Yang Zeros and Dynamical Phases in Quantum Circuits''\par}
\vspace{1em}
Yunfeng Jiang$^{a,c}$, Chang Liu$^{a}$, Yu Wu$^{b}$, and Yang Zhang$^{b,c,d}$\par
\vspace{0.5em}
{\small
$^a$School of Physics \& Shing-Tung Yau Center, Southeast University, Nanjing 211189, P. R. China\\
$^b$Interdisciplinary Center for Theoretical Study, University of Science and Technology of China,
Hefei, Anhui 230026, China\\
$^c$Peng Huanwu Center for Fundamental Theory, Hefei, Anhui 230026, China\\
$^d$Center for High Energy Physics, Peking University, Beijing 100871, People's Republic of China
}
\end{center}
\vspace{1em}

\setcounter{figure}{0}
\setcounter{table}{0}
\setcounter{equation}{0}
\renewcommand{\thefigure}{S\arabic{figure}}
\renewcommand{\thetable}{S\arabic{table}}
\renewcommand{\theequation}{S\arabic{equation}}
\renewcommand{\theHfigure}{supp.\arabic{figure}}
\renewcommand{\theHtable}{supp.\arabic{table}}
\renewcommand{\theHequation}{supp.\arabic{equation}}

\section{Bethe ansatz solution of the brickwork model}
\label{sec:BA}

In this section, we provide details of the Bethe ansatz solution of the brickwork model and the exact calculation of the Loschmidt echo. We consider a model with $L$ qubits and periodic boundary conditions. The Floquet operator of the brickwork circuit reads
\begin{equation}
 \hat{\mathcal{F}}_{\text{bw}}=\mathcal U_{\rm e}(\alpha,\phi)\,
 \mathcal U_o(\alpha,\phi),\qquad
 \mathcal U_{\rm e}=\prod_{j=1}^{L/2}U_{2j,2j+1},
 \quad
 \mathcal U_o=\prod_{j=1}^{L/2}U_{2j-1,2j}.
 \label{sm:eq:brickwork-def}
\end{equation}
In the two-site basis
$\{\ket{00},\ket{01},
\ket{10},\ket{11}\}$, the local gate is the gauge-transformed fSim gate
\begin{equation}
 U_{ij}(\alpha,\phi)=
 \begin{pmatrix}
 1&0&0&0\\
 0&\ee^{-\ri\phi}\cos\alpha&\ri\ee^{-\ri\phi}\sin\alpha&0\\
 0&\ri\ee^{-\ri\phi}\sin\alpha&\ee^{-\ri\phi}\cos\alpha&0\\
 0&0&0&1
 \end{pmatrix}_{ij},
 \label{sm:eq:gate}
\end{equation}
which conserves the number of `$1$'s. In spin-chain terminology, this is the number of magnons. Consequently, $\hat{\mathcal{F}}_{\text{bw}}$ can be diagonalized within sectors of fixed magnon number.

Using the change of variables
\begin{equation}
 \tan\alpha=-\ri\,\frac{\sinh\xi}{\sinh\eta},\qquad
 \ee^{-2\ri\phi}=-\frac{\sinh(\xi-\eta)}{\sinh(\xi+\eta)}\,,
 \label{sm:eq:param}
\end{equation}
the gauge-transformed gate can be identified with the $R$-matrix of the six-vertex model, \emph{i.e.} $U_{ij}(\alpha,\phi)=\check R_{ij}(\xi,\eta)$. More explicitly,
\begin{equation}
 \check R(u)=
 \begin{pmatrix}
 1&0&0&0\\
 0&\dfrac{\sinh\eta}{\sinh(u+\eta)}
   &\dfrac{\sinh u}{\sinh(u+\eta)}&0\\
 0&\dfrac{\sinh u}{\sinh(u+\eta)}
   &\dfrac{\sinh\eta}{\sinh(u+\eta)}&0\\
 0&0&0&1
 \end{pmatrix}.
 \label{sm:eq:R-matrix}
\end{equation}
The anisotropy $\eta$ is kept fixed and suppressed in the notation. The $\check{R}$-matrix satisfies the braid-form Yang--Baxter equation
\begin{equation}
 \check R_{12}(u)\check R_{23}(u+v)\check R_{12}(v)
 =
 \check R_{23}(v)\check R_{12}(u+v)\check R_{23}(u)\,.
 \label{sm:eq:ybe}
\end{equation}
The key observation is that the brickwork Floquet operator can be identified with the product of two staggered row-to-row transfer matrices, which can be diagonalized by the Bethe ansatz. The staggered inhomogeneities are $+\xi/2$ and $-\xi/2$.

In the sector with $M$ magnons, the eigenstates of $\hat{\mathcal{F}}_{\text{bw}}$ can be constructed by the Bethe ansatz and are denoted by $|\mathbf{u}_M\rangle$, where $\mathbf u_M=\{u_1,\ldots,u_M\}$ are the Bethe roots obeying the following Bethe ansatz equations:
\begin{equation}
 \left[
 \frac{\sinh(u_j-\xi/2+\eta)}{\sinh(u_j-\xi/2)}
 \frac{\sinh(u_j+\xi/2+\eta)}{\sinh(u_j+\xi/2)}
 \right]^{L/2}
 =
 \prod_{\substack{k=1\\ k\ne j}}^M
 \frac{\sinh(u_k-u_j+\eta)}
      {\sinh(u_k-u_j-\eta)} ,
 \qquad j=1,\ldots,M .
 \label{sm:eq:bae}
\end{equation}
The Floquet eigenvalue $\tau(\mathbf u_M)$ is given by
\begin{equation}
 \tau(\mathbf u_M)
 =
 \prod_{k=1}^M
 \frac{\sinh(u_k-\xi/2+\eta)\sinh(u_k+\xi/2)}
      {\sinh(u_k+\xi/2+\eta)\sinh(u_k-\xi/2)} \,.
 \label{sm:eq:floquet-eig}
\end{equation}
%Singular solutions and root-of-unity degeneracies require the usual limiting prescription; they are not used as independent states in the numerical data reported below.

For an initial state $\ket{\Psi}$ in a fixed magnon sector, the Loschmidt amplitude is
\begin{equation}
 \mathcal D_\Psi(n)=\bra{\Psi}\hat{\mathcal{F}}_{\text{bw}}^n\ket{\Psi}
 =
 \sum_{\text{sol}}
 w_\Psi(\mathbf u_M)\,\tau(\mathbf u_M)^n\,,
 \label{sm:eq:loschmidt-ba}
\end{equation}
where the sum runs over all physical solutions.
The weight $w_\Psi(\mathbf u_M)=|\braket{\Psi}{\mathbf u_M}|^2/
\braket{\mathbf u_M}{\mathbf u_M}$ is related to the overlap of the initial state $|\Psi\rangle$ with the Bethe state. The norm of the Bethe states can be written in terms of the Gaudin determinant~\cite{Gaudin1983Bethe,korepin1982BEF}. For certain initial states, the overlap $\braket{\Psi}{\mathbf u_M}$ can be computed efficiently by determinant formulae. We consider two such families of initial states. The first consists of domain-wall states, whose overlap formula can be found in Refs.~\cite{MosselCaux2010DomainWall,Slavnov1989ScalarProducts,Izergin1987partition,Izergin1992,Kostov2012IPBS}. The second family comprises integrable boundary states. Determinant formulas are available for the N\'eel and dimer states~\cite{Brockmann2014NeelXXZ,Pozsgay2014ProductOverlaps,JiangPozsgay2020ExactOverlaps} and for the spin-chain crosscap state~\cite{CaetanoKomatsu2022Crosscap}.\par

The product states used in the main text are
\begin{align}
 \ket{\mathrm{DW}_M}
 &=
 \ket{\underbrace{1\cdots1}_{M}
       \underbrace{0\cdots 0}_{L-M}},
 \nonumber\\
 \ket{\mathrm{N\acute eel}}
 &=
 \ket{0101\cdots},
 \nonumber\\
 \ket{\mathrm{Dimer}}
 &\propto
 \prod_{j=1}^{L/2}
 \bigl(\ket{1_{2j-1}0_{2j}}
      -\ket{0_{2j-1}1_{2j}}\bigr),
 \nonumber\\
 \ket{\mathrm{Crosscap}}
 &\propto
 \prod_{j=1}^{L/2}
 \bigl(\ket{0_j0_{j+L/2}}
      +\ket{1_j1_{j+L/2}}\bigr),
 \label{sm:eq:initial-states}
\end{align}
with the appropriate projection to a fixed magnetization sector whenever needed. Overall normalization factors do not affect the zeros.\par

To obtain all physical solutions of the Bethe ansatz equations, we can apply the rational $Q$-system approach developed in \cite{Bajnok2020AG,Hou:2023jkr}. To perform the sum over all states, we use the algebro-geometric method developed in \cite{Jiang2018AG,Jacobsen2019AG,Bajnok2020AG}. A detailed introduction to these approaches in the isotropic case can be found in \cite{Hutsalyuk2024sawAG}, while the generalization to the anisotropic case, including a more detailed study at roots of unity, will be given in \cite{ToAppear}. 

\section{Results for small sizes}
\label{sec:small-sizes}

Strictly speaking, the analysis of the condensation curve using the BKW theorem holds in the $n\to\infty$ limit. However, the distribution patterns and the abrupt reorganization can already be seen clearly even for relatively small $n$, which facilitates experimental verification. In this section we display representative examples to demonstrate this point. We consider the domain-wall state with $L=8$ in the $M=2$ sector,
\begin{equation}
 \ket{\mathrm{DW}_2}=\ket{11000000}.
\end{equation}
At fixed anisotropy $q=q_0$, we write
\begin{equation}
 \mathcal D_{8|q_0}^{(2)}(n;x)
 =
 \bra{\mathrm{DW}_2}\mathcal U(q_0,x)^n\ket{\mathrm{DW}_2}.
 \label{sm:eq:D-L-q}
\end{equation}
The GPLY zeros are the zeros of the numerator after cancellation of common factors with the denominator.

\begin{figure}[h!]
\centering
\setlength{\abovecaptionskip}{2pt}
\setlength{\belowcaptionskip}{0pt}
\begin{minipage}{0.48\textwidth}
 \centering
 \includegraphics[width=0.62\linewidth]{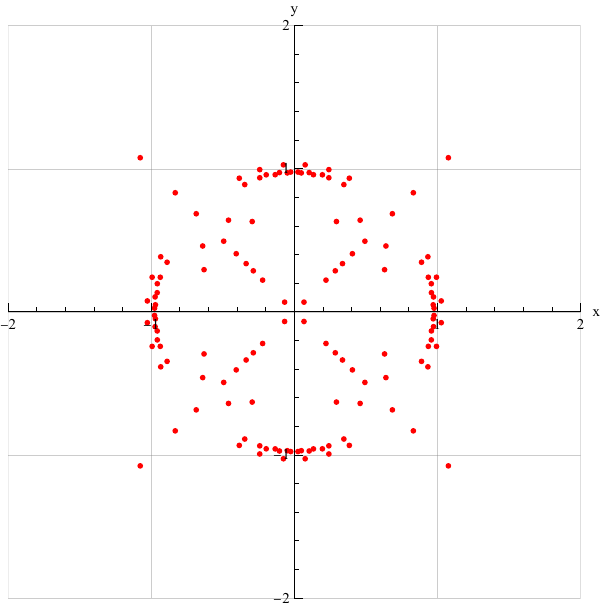}
 \caption{GPLY zeros of $\mathcal D_{8|2}^{(2)}(10;x)$ for the domain-wall initial state for $q_0=2$.}
 \label{sm:fig:q2n10}
\end{minipage}
\hfill
\begin{minipage}{0.48\textwidth}
 \centering
 \includegraphics[width=0.62\linewidth]{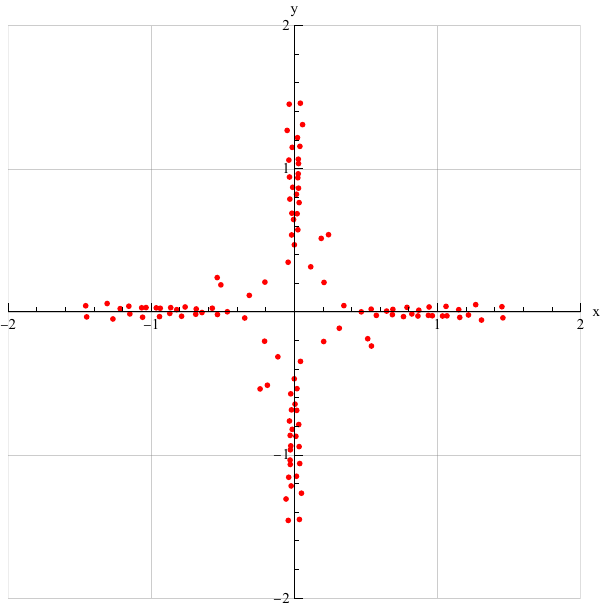}
 \caption{GPLY zeros of $\mathcal D_{8|3/5+4\ri/5}^{(2)}(10;x)$ for the domain-wall initial state for $q_0=3/5+4\ri/5$.}
 \label{sm:fig:q3-4in10}
\end{minipage}
\end{figure}

In the massive regime, a useful benchmark is $q_0=2$. For $n=10$ one finds
\begin{equation}
 \mathcal D_{8|2}^{(2)}(10;x)=
 \frac{P_{8|2}^{(2)}(x)}{(1-4x^4)^{38}},
 \label{sm:eq:massive-rational}
\end{equation}
The zeros of $P_{8|2}^{(2)}(x)$ is plotted in Fig.\ref{sm:fig:q2n10}. In the massless regime we use
$q_0=\frac{3}{5}+\frac{4\ri}{5}$. For the same system size and time,
\begin{equation}
 \mathcal D_{8|3/5+4\ri/5}^{(2)}(10;x)=
 \frac{P_{8|3/5+4\ri/5}^{(2)}(x)}
 {((3+4\ri)x^4-(3-4\ri))^{38}} .
 \label{sm:eq:massless-rational}
\end{equation}
The zeros of $P_{8|3/5+4\ri/5}^{(2)}(x)$ is plotted in Fig.\ref{sm:fig:q3-4in10}. \par

The massive and massless finite-size examples below already show the qualitative distinction used in the main text: in the massive regime the zeros organize into compact closed curves, whereas in the massless regime they form extended radial structures. This contrast is sharpened by increasing $L$ and $n$. The plots show the zeros for two different times.

\section{Direct calculation of the Loschmidt amplitude for general circuits}
\label{app:direct-calculation}

For a general Floquet operator such as the $\hat{\mathcal{F}}_{\text{II},\text{III}}$ which are not solvable by Bethe ansatz, we take a different approach and evaluate the action of $\hat{\mathcal{F}}$ on quantum states directly. Since the two-site gate preserves the number of magnons, we can calculate the Loschmidt amplitude within a fixed $M$-magnon sector. We denote the $M$-magnon basis by $\mathcal B_M$, whose dimension is $\binom{L}{M}$. 
The gauge-transformed fSim gate can be written in the following form
\begin{equation}
 U_{ij}(q,x)=
 \begin{pmatrix}
 1&0&0&0\\
 0& c(q,x)
   & b(q,x)&0\\
 0& b(q,x)
   &c(q,x)&0\\
 0&0&0&1
 \end{pmatrix}_{ij}.
\end{equation}
Locally on sites $i$ and $j$, the gate acts as
\begin{align}
\label{eq:actU}
 U_{ij}\ket{00}&=\ket{00},
 &U_{ij}\ket{11}&=\ket{11},\nonumber\\
 U_{ij}\ket{01}
 &=c\ket{01}+b\ket{10},
 &U_{ij}\ket{10}
 &=b\ket{01}+c\ket{10}.
\end{align}
where $b$ and $c$ are given by
\begin{equation}
 b=b(q,x)=\frac{q(1-x^4)}{1-q^2x^4},
 \qquad
 c=c(q,x)=\frac{(1-q^2)x^2}{1-q^2x^4}.
\end{equation}
After $m$ complete Floquet periods, we write the resulting state as
\begin{equation}
 \ket{\Psi^{(m)}}=
 \sum_{\boldsymbol s\in\mathcal B_M}
 A_{\boldsymbol s}^{(m)}(b,c)\ket{\boldsymbol s},
 \qquad
 A_{\boldsymbol s}^{(0)}=\braket{\boldsymbol s}{\Psi}.
\end{equation}
The coefficients $A_{\boldsymbol s}^{(m)}(b,c)$ are polynomials in $b$ and $c$. In our implementation, we store only the coefficient and the powers of $b$ and $c$ to save memory. 
To obtain the state for $m+1$ periods, we apply the two-site rule \eqref{eq:actU} above
successively along the ordered list of bonds defining the Floquet operator. This can be implemented efficiently using sparse matrix products.
Repeating the same procedure generates the sequence
\begin{equation}
 \ket{\Psi^{(0)}}\longrightarrow
 \ket{\Psi^{(1)}}\longrightarrow\cdots\longrightarrow
 \ket{\Psi^{(n)}}.
\end{equation}
The large-$n$ expression is therefore obtained by direct iteration. 

As an example, we consider the action of $\hat{\mathcal{F}}_{\text{II}}$ on the $L=8$, $M=2$ domain-wall state. The first layer acts trivially,
$\mathcal U_o\ket{\mathrm{DW}_2}=\ket{\mathrm{DW}_2}$. The range-two layer
$\mathcal V_2$ is then applied in the ordered sequence
$U_{1,3},U_{2,4},\ldots,U_{8,2}$. For example, the first gate gives
\begin{align}
 U_{1,3}\ket{\mathrm{DW}_2}
 &=
 c\ket{11000000} +b\ket{01100000}.
\end{align}
The remaining gates are applied by the same local rule \eqref{eq:actU}. Completing
the ordered sequence yields
\begin{equation}
 \hat{\mathcal F}_{\mathrm{II}}\ket{\mathrm{DW}_2}
 =
 \left(c^4+2b^4c^2+b^8\right)\ket{\mathrm{DW}_2}
 +\ket{\Psi_{\perp}^{(1)}},
 \qquad
 \braket{\mathrm{DW}_2}{\Psi_{\perp}^{(1)}}=0,
\end{equation}
where $\ket{\Psi_{\perp}^{(1)}}$ collects all other configurations in the
$M=2$ sector. Hence the one-period Loschmidt amplitude is
\begin{equation}
 \mathcal D_{\mathrm{DW}_2}(1;b,c)
 =
 \bra{\mathrm{DW}_2}\hat{\mathcal F}_{\mathrm{II}}
 \ket{\mathrm{DW}_2}
 =
 c^4+2b^4c^2+b^8.
\end{equation}
To reach larger depths, the complete state, including
$\ket{\Psi_{\perp}^{(1)}}$, is used as the input for the next Floquet period.
The projection onto the initial state is performed only after the desired
number of periods.

After depth $n$ is reached, the Loschmidt amplitude is obtained by
projecting the final state back onto the initial state,
\begin{equation}
 \mathcal D_\Psi(n;b,c)=\braket{\Psi}{\Psi^{(n)}}
 =\sum
 \braket{\Psi}{\boldsymbol s}A_{\boldsymbol s}^{(n)}(b,c).
\end{equation}
For a domain-wall state, this simply selects the polynomial amplitude of the initial domain-wall configuration. Finally, at fixed $q$ we cancel
common numerator--denominator factors exactly and obtain the GPLY zeros from the reduced numerator.

\end{document}